\documentclass[superscriptaddress, reprint, showpacs, english, aip, jcp]{revtex4-1}

\usepackage{amsmath} 
\usepackage{overpic} 
\usepackage{graphicx}
\usepackage{hyperref} 
\usepackage{tabularx} 
\usepackage{verbatim}

\newcolumntype{L}[1]{>{\raggedright\arraybackslash}p{#1}} 
\newcolumntype{C}[1]{>{\centering\arraybackslash}p{#1}} 
\newcolumntype{R}[1]{>{\raggedleft\arraybackslash}p{#1}} 

\newcommand{\kb}{k_{\mathrm{B}}}
\newcommand{\upperef}{\omega_\mathrm{u}}
\newcommand{\loweref}{\omega_\mathrm{l}}
\newcommand{\lvr}{\langle v\rangle}

\begin{document}

\title{Optimizing the Performance of the Entropic Splitter for Particle Separation}

\date{\today}

\author{T. \surname{Motz}}
\altaffiliation{New address: Institut f\"ur Theoretische Physik, Universit\"at Ulm, Albert-Einstein-Allee 11, 89069 Ulm, Germany}
\email{thomas.motz@uni-ulm.de}
\affiliation{Institute  of Physics, University of Augsburg,
Universit\"atsstrasse 1, D-86135,  Augsburg, Germany}

\author{G. \surname{Schmid}}
\affiliation{Institute  of Physics, University of Augsburg,
Universit\"atsstrasse 1, D-86135,  Augsburg, Germany}

\author{P. H\"anggi}
\affiliation{Institute  of Physics, University of Augsburg,
Universit\"atsstrasse 1, D-86135,  Augsburg, Germany}
\affiliation{Nanosystems Initiative Munich, Schellingstr, 4, D-80799 M\"unchen, Germany}
\author{D. \surname{Reguera}}
\affiliation{Departament de F\'{i}sica Fonamental, Facultat de F\'{i}sica, Universitat de Barcelona, Mart\'{i} i Franqu\`{e}s 1, E-08028 Barcelona, Spain}
\author{J. M. \surname{Rub\'{i}}}
\affiliation{Departament de F\'{i}sica Fonamental, Facultat de F\'{i}sica, Universitat de Barcelona, Mart\'{i} i Franqu\`{e}s 1, E-08028 Barcelona, Spain}

\keywords{entropic splitter; confined diffusion; particle separation}
\pacs{05.40.-a, 02.50.Ey, 05.10.Gg, 05.60.-k}

\begin{abstract}
Recently, it has been shown that entropy can be used to sort Brownian particles according to their size. In particular, a combination of a static and a time-dependent force applied on differently sized particles which are confined in an asymmetric periodic structure can be used to separate them efficiently, by forcing them to move in opposite directions. In this paper, we investigate the optimization of the performance of the ``entropic splitter''. Specifically, the splitting mechanism and how it depends on the geometry of the channel, and the frequency and strength of the periodic forcing is analyzed. Using numerical simulations, we demonstrate that a very efficient and fast separation with a practically 100\% purity can be achieved by a proper optimization of the control variables. The results of this work could be useful for a more efficient separation of dispersed phases such as DNA fragments or colloids dependent on their size.



\end{abstract}

\maketitle


\section*{I. Introduction}

In many natural systems and industrial applications, matter consists not of pure substances but of mixtures of different components. Examples include polydispersed colloids and polymers, red and white blood cells, or healthy and cancerous cells \cite{Li_Peng,Cristofanilli,Geislinger}. 
The capability of sorting these different constituents to achieve pure substances is thus a crucial challenge with very important technological applications in industry, nanotechnology, and biology. These components, dispersed in a liquid phase, diffuse in and eventually are convected by the presence of external and internal forces. Confinement can also play a very significant role on a wide range of 
systems and scales starting on 
the nanoscale with ions and molecules up to 
objects in the micrometer range like DNA or red blood cells \cite{Hille, Bakshi}. 

To find a way to sort these objects, one is always confronted with an interplay of different attributes like charge, mass, and size 
that lead to differential responses to the application of an external field.
A separation dependent on 
particles' mass and size is usually carried out 
by centrifugation \cite{Harrison}, whereas electrophoresis 
combines diffusion through a gel with an external electric field to sort out particles of different sizes and charges~\cite{Slater, Volkmuth, Dorfman}. A purely size-dependent separation in its simplest form is realized by a sieve or 
porous materials \cite{Corma, Haul}. 
The current techniques proposed for particle separation 
rely on forcing the particles to move at different velocities but in the same direction. Other separation techniques based on size-dependent hydrodynamical long-range interactions \cite{Geislinger, Loutherback} or ratcheting of particles moving in asymmetric arrays or channels have been suggested \cite{Duke, Eichhorn_Reimann, Reimann2012, Kettner, Mueller, Haenggi_2009, Davis_John, Burada_chemphyschem, Rubi_2010, Rubi_2013}.

Recently, a novel splitting mechanism 
based on entropic rectification 
was presented 
and suggested as a device which can be used for efficient particle sorting in a time continuous mode \cite{Schmid}. 
The essence of this mechanism relies on the use of an asymmetric and periodic channel to confine spherical particles of different radii, employing a combination of different forces to drive their motion. An unbiased time-periodic force acting on particles confined in an asymmetric and periodic channel leads to a rectification of the noisy motion and to a net-drift along a preferential direction (i.e. the one with the least steep walls). 
This effect becomes stronger with increasing particle radius. Applying an additional small static force in the opposite direction, one is able to set up a configuration where particles below a certain radius travel to the opposite direction as the larger particles do. If one varies the parameters of the channel or the external forces, the critical radius that dictates in which direction particles move is under perfect control with this set up. 
That is the essential concept of the entropic splitter that was introduced in Ref. \cite{Schmid} using a specific channel geometry and set of parameters. 

Whereas the entropic splitter seems to be a very promising device to sort particles of different sizes, there are still open questions about its efficiency and control that have to be addressed before proceeding with an experimental implementation. To study how to optimize the operation of this potential device is precisely the aim of this article. More specifically, we are going to analyze how the different parameters affect the performance of the entropic splitting mechanism. We will discuss how a smartly choosen set of parameters can be used to optimize the efficiency of entropic rectification and separation, and verify its feasibility using specific examples. 

One of the most important factors influencing the performance of the entropic splitter is the shape of the periodic confining channel. In Ref. \cite{Schmid}, a simple saw-tooth profile with two different and relatively small slopes was used to facilitate the applicability of the Fick-Jacobs (FJ) approximation \cite{Jacobs, Zwanzig, Reguera, Burada}. However, since (entropic) rectification relies on asymmetry, a channel structure as that shown in Fig.~\ref{Geometrie} is the one offering the strongest rectification, as it was verified for point particles in recent publications \cite{Dagdug_2012, *Marchesoni_2009, *Borromeo_2011, *Marchesoni_2010}. For the sake of concreteness, we will focus on this structure which maximizes rectification. The small drawback is that, given the steepness of the vertical wall, the FJ approximation is expected to be not very accurate \cite{Burada}. Accordingly, mostly numerical studies will be performed.

\section*{II. Model}

The dynamics of a Brownian particle with an external static force $f$ and an additional time-dependent force $F(t)$ acting along the $x$-axis can be described by the overdamped Langevin equation \cite{Schmid}
\begin{equation}
\gamma\frac{\mathrm{d}\vec{r}}{\mathrm{d}t}=\mathbf(f+F(t)\mathbf)\vec{e}_x+\sqrt{2\gamma\kb T}\vec{\xi}(t)\;.
\label{Langevin_full}
\end{equation}
Since we assume hard spheres $\gamma$ is given by the Stokes' friction, $\vec{r}$ is the particle's position in the 2D channel, $\vec{e}_x$ the unit vector in $x$-direction and $\kb T$ represents the thermal energy of the heat bath. The Gaussian random force $\vec{\xi}(t)$ with zero mean is uncorrelated in time and therefore obeys the condition for white noise given by \mbox{$\langle \xi_i(t)\,\xi_j(t')\rangle=2\delta_{ij}\delta(t-t')$} with $i,j=x,y$. The overdamped limit will be considered, corresponding to $t\gg m/\gamma$. Under this conditions, effects caused by the particle's inertia can safely be neglected \cite{Kramers, Purcell}. $F(t)$ is a periodic square wave with frequency $\omega$ and amplitude $A$ explicitly given by $F(t)=A\,\mathrm{sgn}[\sin(\omega t)]$,  and acts along $\vec{e}_x$ according to Eq.~(\ref{Langevin_full}).

\begin{figure}
\includegraphics[width=8cm]{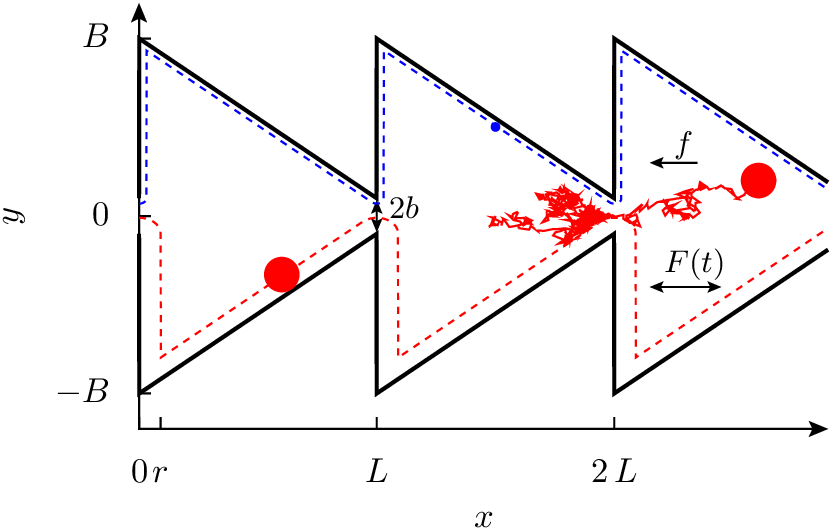}
\caption{The two-dimensional geometry of the channel with an exemplary trajectory of a diffusing particle with $r=0.9\,b$. The channel has a periodicity of $L$, a slope $m=0.9$ and a bottlenecks' half-width $b=0.1\,L$, whereas the total width is given by $B=b+mL$. The dashed lines represent the effective boundaries given by $\upperef (x)$ and $\loweref (x)$ that confine the region accessible for the center of a spherical particle with radius $r=0.9\,b$ (red) and $r=0.3\,b$ (blue). 
}
\label{Geometrie}
\end{figure}

The Langevin equation has to be solved with reflecting boundary conditions at the channel walls. The 2D channel with period length $L$ and bottleneck width $2b$ is shown in Fig.~\ref{Geometrie} with an exemplary trajectory of a diffusing particle. Taking also the total width $2B$ and so the slope $m=(B-b)/L$ into account, the channel geometry is fully defined by the equation for the upper boundary $y_\mathrm{u}(x)$ 
\[y_\mathrm{u}(x)=b+m(L-\bar{x})\;,\]
where 
$\bar{x}$ is given by the modulo function $\bar{x}=x\mod L$ and causes a periodic structure. Due to the symmetry, the lower boundary $y_\mathrm{l}(x)$ is given by $y_\mathrm{l}(x)=-y_\mathrm{u}(x)$.

When implementing hard-spheres, one has to modify the accessible space for the center of the particles by introducing an effective boundary. This can be constructed by drawing the center's trajectory of a particle that moves along the boundary $y_\mathrm{u}(x)$, creating circles in the vicinity of the bottlenecks starting at the positions \mbox{$L_\mathrm{p}=L-r m/\sqrt{1+m^2}$} and parallel lines with a vertical distance of  $h=r\sqrt{1+m^2}$ along the straight parts. Translated into formulas, this argumentation leads to the upper effective boundary $\upperef(x)$:
\begin{equation}
\upperef(x)=
\begin{cases} 
b-\sqrt{r^2-\bar{x}^2} & 0\leq\bar{x}<r\\ 
b+m\,(L-\bar{x})-r\,\sqrt{1+m^2} & r<\bar{x}\leq L_\mathrm{p}\\ 
b-\sqrt{r^2-(\bar{x}-L)^2} & L_\mathrm{p}<\bar{x}\leq L
\end{cases}\,.
\label{effbound} 
\end{equation}
The lower effective boundary 
is given by $\upperef(x)=-\loweref(x)$. 
The result is shown by the dashed lines in Fig. \ref{Geometrie} where the smaller accessible space for the centers of spherical particles compared to point particles, and its dependency on particle size, is visible.

In order to achieve a dimensionless description, all quantities that occur in the Langevin Eq.~(\ref{Langevin_full}) will be scaled in terms of the characteristic period length $L$ and diffusion time $\tau=L^2/D_b$, where \mbox{$D_b=\kb T/\gamma_b$} and \mbox{$\gamma_b=6\pi\eta b$} are the diffusion constant and Stokes' friction of a spherical particle with a reference radius $r=b$, respectively. In terms of these characteristic parameters, we have:
\begin{equation}
\widehat{\omega}_\mathrm{u}(x)=\frac{\upperef(x)}{L}\phantom{.........}\widehat{x}=\frac{x}{L} \phantom{.........} \hat{t}=\frac{t}{\tau} \phantom{.........} \widehat{\omega}=\omega\,\tau\;,
\label{scaling}
\end{equation}

Moreover, to simulate DNA in an external electrical field we shall assume the force to depend linearly on the radius $r$:
\begin{equation}
f = f_0\frac{r}{b}\phantom{...........}A = F_0\frac{r}{b}\;.
\end{equation} 
With these transformations 
the scaled version of the Langevin equation reads
\begin{equation}
\frac{\mathrm{d}\vec{r}}{\mathrm{d}t}=\mathbf(f_0+F_0\,\mathrm{sgn}[\sin(\omega t)]\mathbf)\vec{e}_x+\sqrt{2 b/r}\vec{\xi}(t)\;,
\label{Langevin_scaled}
\end{equation}
where for the sake of simplicity the hat-symbols have been omitted.

To complete the model description, strong viscous dynamics in the overdamped limit combined with a dilute density that frustrate hydrodynamic particle-wall interactions and particle-particle interactions are assumed \cite{Martens_hydro,  Happel, Maxey}. This setting, for example, mimics the situation in blood with cancerous cells where $\sim 1$ tumor cell exists within one mililiter of blood which has a low Reynolds number and basically consists of plasma and red and white blood cells which are much smaller than tumor cells \cite{Chen, Davis_John, Parichehreh, Hur}. For studies that take hydrodynamic interactions into account see Refs. \cite{Martens_hydro, Martens_hydro_2}.

An alternative, and approximate, description of this system can be performed in terms of the corresponding Fick-Jacobs equation including an entropic potential $-TS(x)=-\ln[2\,w(x)]$, which effectively accounts for the effects of confinement \cite{Reguera, Reguera2006, Jacobs, Zwanzig}. Due to the presence of a steep wall in the channel (see Fig.~\ref{Geometrie}) the FJ is expected to be not very accurate even for very small bias. Nevertheless, the shape of the entropic barrier and its height provide insightful information on the transport behavior in this system. 

There are two parameters of the channel that have an obvious influence on the height of the entropic potential and accordingly on the efficiency of the entropic rectification and splitting. One is the width of the bottleneck $b$. Obviously, the smaller the bottleneck width, the higher the entropic barrier and the rectification. The second parameter is slope $m$ of the wall. An increase of $m$ will lead to larger channel's  half-width $B=b+mL$, and accordingly also to higher barriers and stronger rectification. However, beyond a typical value of the slope around $m\sim 10$,  this enhancement is no longer significant since for very steep walls and large total widths $2B$ the particle distribution does not spread over the whole channel-width and therefore a further increase of the space has no impact to the net-drift of the particles. This was corroborated in our simulations. Accordingly, we will fix the values of $b$ and $m$, and explore how the other parameters affect the rectification efficiency.

\section*{III. Non-linear Mobility}

Macroscopic transport quantities like the mobility are calculated by averaging over an ensemble of $10^4$ trajectories that are simulated  based on Eq.~(\ref{Langevin_scaled}) via a Stochastic Euler procedure \cite{Kloeden}. 
In the following, we will analyze the influence of different parameters on the rectification and splitting efficiency.
First, we study the system's response behavior by applying a static force along the principal axis of the channel. This causes a mean velocity which is closely related to the non-linear mobility $\mu$, defined as
\begin{equation}
\mu=\lim_{t\to\infty}\frac{\lvr}{f_0}\;,
\label{mobility}
\end{equation}
where $f_0$ represents the external static force and $\lvr$ the mean velocity calculated from an ensemble of simulated trajectories.

\begin{figure}
\includegraphics[width=8cm]{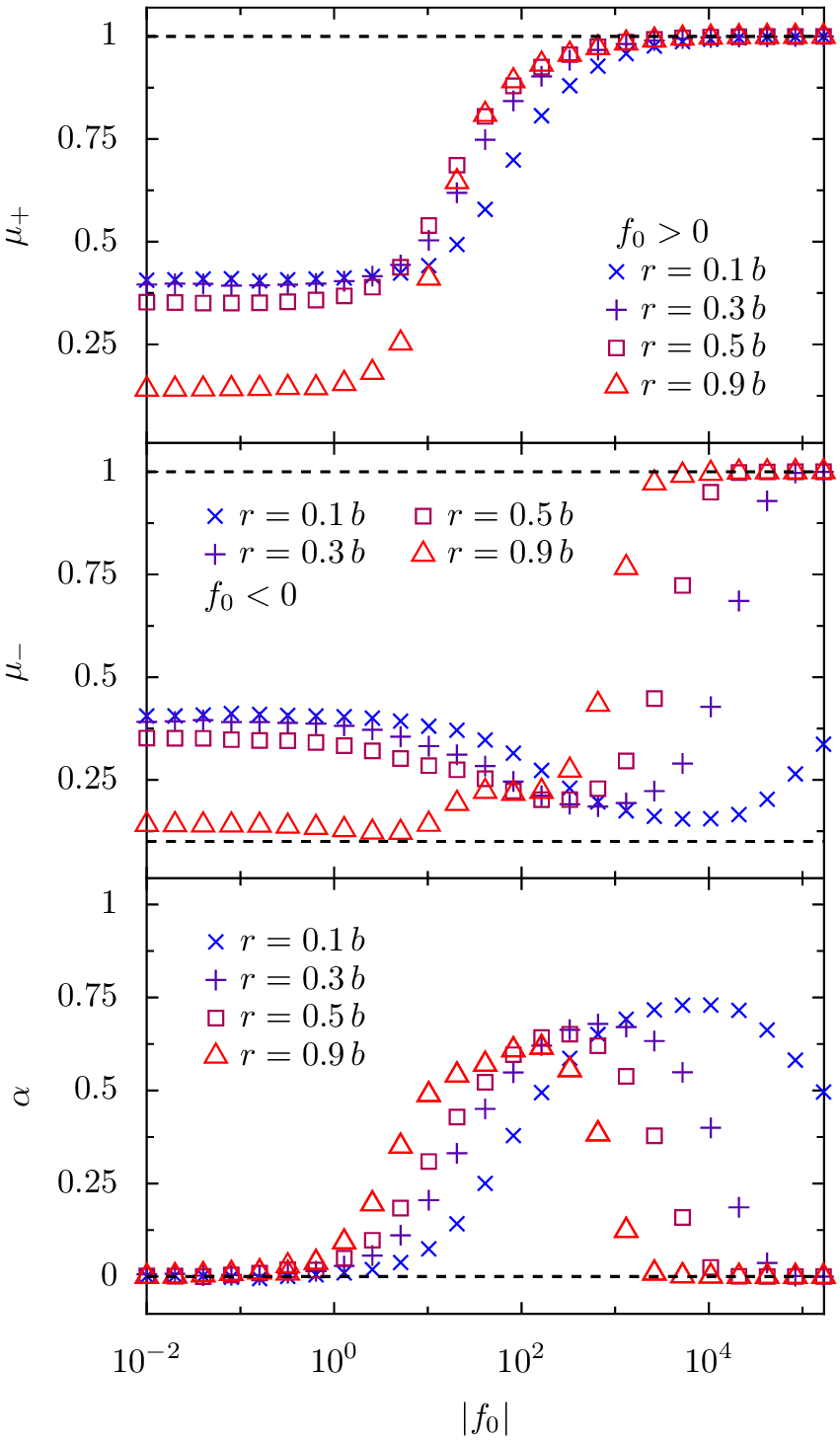}
\caption{The non-linear mobility $\mu=\lim_{t\to\infty}\langle\dot{x}\rangle/f_0$ of spherical particles with different radii $r$ moving in the channel shown in Fig.~\ref{Geometrie}. The upper plot shows $\mu$ for an external static bias $f_0$ in positive direction, whereas $f_0$ is negative in the middle plot. The dashed lines show the analytic limits for point-like particles, namely $\lim_{f_0\to\infty}\mu=1$ for a positive $f_0$ and $\lim_{f_0\to -\infty}\mu=b/B$ for a negative $f_0$ (lower line in the middle plot). The lower plot shows the rectification coefficient $\alpha$, characterized by a fast decrease of $\alpha$ for large particle radii. An ensemble of $10^4$ non-interacting particles was simulated until an accuracy of $10^{-3}$ for $\mu$ was achieved.}
\label{muoverf_R_vgl_app}
\end{figure}

Fig. \ref{muoverf_R_vgl_app} plots the mobility as a function of the static bias, in the absence of a periodic force. The upper plot shows $\mu_+$ that represents the mobility for an $f_0$ that is positive and points to the right in Fig.~\ref{Geometrie}. 
For small forces $f_0 < 1$,  $\mu_+$ is almost constant with a different value for each particle radius $r$. As $f_0$ gets stronger, $\mu$ increases and eventually converges to $\mu_+=1$ for all particle radii if $f_0$ is large enough. The inflection point of the mobility curve for any given radius roughly coincides with the vanishing of the entropic barrier, which occurs at values of the external force equal to the force at the inflection point of the entropic potential, given by $f_\mathrm{infl}=\frac{m b}{r(b- r/\sqrt{1+m^2})}$. For \mbox{$f_0>10^3$}, the forcing is so strong that all particles move in a lane in the center of the channel which has a width of $2b$. Therefore, the particle's motion is not disturbed by the channel walls for these values of $f_0$, leading to $\mu_+=1$.

The middle plot shows $\mu_-$ for $f_0<0$, which is equivalent to an $f_0$ pointing to the left in Fig.~\ref{Geometrie}. For small $|f_0|$s,  the mobility
$\mu_-$ is constant with a value that depends on $r$. As $|f_0|$ is increased, the mobility shows a non-monotonic behavior which strongly depends on $r$. Interestingly, for very large $|f_0|$s the mobility of all particles converges back to one. This is in contrast to the behavior of point particles ($r=0$) in the same channel, where $\mu_-$ converges to a limiting value of $\mu_-=b/B$ for large negative forces. That implies that the particles are uniformly distributed along the $y$-axis. The convergence to $\mu_-=1$ we found here is an effect that has its  origin in the rounded bottlenecks of the effective confinement $\upperef(x)$ and $\loweref(x)$ described by Eq.~(\ref{effbound}).  When the effective shape around the bottleneck is rounded,  a particle moving in the negative direction and touching the confinement in a range of \mbox{$b-r\leq |y|\leq b$} will be eventually guided into the bottleneck even at very large negative forcings, and so contributes to the transport.  In contrast, a point particle that hits a vertical wall perpendicular to the force's direction will get stuck against the wall at very large forces, and will not be able to pass through the bottleneck. 
This extreme sensitivity of the limiting value of the mobility to the slope near the bottleneck was already found for point particles by Dagdug \textit{et al.} in Ref.~\cite{Dagdug_2012} by investigating periodic channels where the compartments first are separated by vertical boundaries, leading to a monotonic decrease of $\mu$ and, second, by boundaries with finite slopes what leads to the behavior found here for hard spherical particles.

The strong dependency on the particle radius of the typical forces required for recovering $\mu_-=1$ can be understood using a simple estimate based on diffusion. Essentially all particles of a given radius $r$ will pass through the bottleneck if the distance travelled by diffusion in the vertical direction, $y=\sqrt{2\,b/r\,t}$, during the time required to cross one period of the channel, $t=1/\lvr$, is smaller than the bottleneck width $b$. Assuming that the velocity is roughly proportional to the force $\lvr \sim f_0$, the above criterion leads to a simple estimate 
$f_b=\frac{2}{r b}$
of the force, in reduced units, beyond which the particles diffuse in the vertical direction a distance smaller than the bottleneck radius $b$, and thus are expected to be focused through the channel. 
This simple argument explains why $\mu_-$ starts to converge to one at smaller forces for large particles. 
There is also another striking feature in the behavior of $\mu_-$ for $r=0.9b$. A plateau in the mobility is visible for intermediate values of the force around $10^2$. This plateau seems to be associated to the existence of a region of negative forces where the height of the effective entropic barrier becomes nearly constant. 
The complex nonmonotonic behavior of the mobility for $f_0<0$ already suggests that making an adequate choice of the external force's value for an optimal particle separation is not a trivial issue.

To get an idea of the dynamics for an applied oscillating force in the adiabatic regime, the lower plot in Fig.~\ref{muoverf_R_vgl_app} shows the rectification coefficient defined as \cite{Kosinska_2008, Schmid_2009} 
\begin{equation}
\alpha=\frac{\mu_+-\mu_-}{\mu_+ +\mu_-}\;.
\end{equation}  
As expected, $\alpha$ initially rises with $|f_0|$ for all radii but more intensively the larger the $r$ is. Accordingly, there is entropic rectification and larger particles drift on average to the right at larger velocities than smaller particles. However, 
for $|f_0|\geq 10^2$ the values of $\alpha$ start to decrease beyond a critical force that depends on the particle radius $r$, and in fact for large $r$ drop below the ones for small $r$. This is caused by the accumulation of the particles along the channel's principal axis for large negative $f_0$ that progressively destroys the rectification effect.  As described before, the forces where this accumulation occurs are smaller the larger the particle radius is. Thus, at very large amplitudes of the oscillating force, smaller particles are rectified more efficiently than larger particles.
Eventually, for large $f_0$ all particles are accumulated along the center for both negative and positive forces, the dynamics in both directions are equal and therefore $\alpha$ vanishes. 
The maximum in the rectification coefficient $\alpha$ roughly coincides with the minimum in the negative mobility, and is thus controlled by the onset of the channeling effect discussed above. 

The previous results suggest that amplitudes $F_0$ of the oscillating force in the range $10\leq F_0\leq 10^2$ are considered to be adequate for an optimal particle separation since the accumulation effects for larger forces make it difficult to adjust an additional static force in order to let two different kinds of particles move in opposite directions.

\section*{IV. Beyond the Adiabatic Limit}\label{Beyond_adiabatic}

\begin{figure}
\begin{center}
\includegraphics[width=8cm]{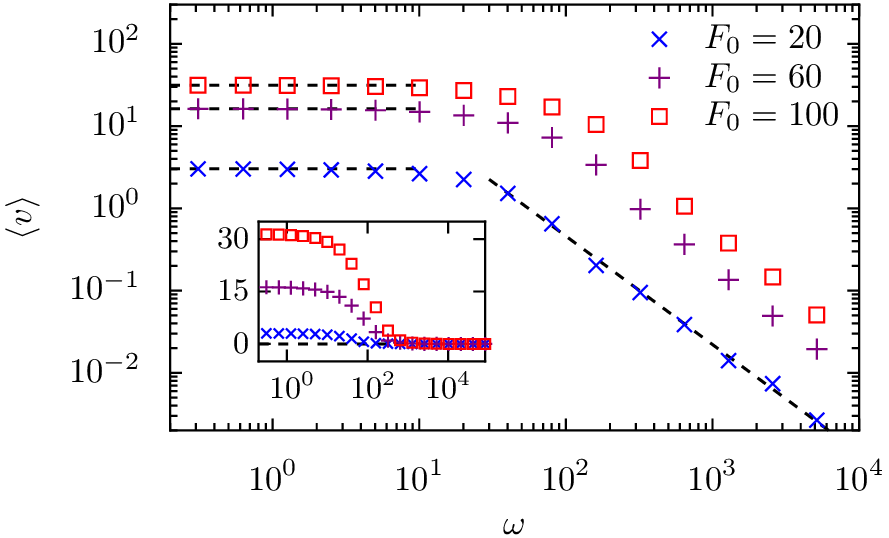}
\end{center}
\caption{The mean velocity $\lvr$ of particles with $r=0.3\,b$ moving in the channel of Fig.~\ref{Geometrie} versus the frequency $\omega$ for different amplitudes $F_0$ of the oscillating force $F(t)$ on a double logarithmic scale, and on a logarithmic scale shown in the inset. The dashed lines represent the adiabatic limit calculated as the arithmetic mean \mbox{$\lvr=(\langle v\rangle_++\langle v\rangle_-)/2$} of simulation results obtained with static forces with the particular values $F_0$ and $-F_0$ leading to the velocities $\langle v\rangle_+$ and $\langle v\rangle_-$, respectively. 
The dashed line for $\omega > 10^2$ is a fitted result according to the law $\omega^a$.}
\label{v_over_freq}
\end{figure}

Beside the amplitude $F_0$, the frequency $\omega$ is the second variable of the oscillating force $F(t)$ that needs to be fixed. Let us now analyze how the velocity of particles of a given size depends on the frequency of the forcing beyond the adiabatic limit. Fig.~\ref{v_over_freq} shows the mean velocity $\lvr$ caused by an oscillating drive $F(t)$ versus $\omega$ for different amplitudes $F_0$ on a double-logarithmic scale and on a logarithmic one (inset). For small values of $\omega$ the mean velocity is almost constant and coincides with the expected value in the adiabatic limit given by \mbox{$\langle v\rangle=(\langle v\rangle_++\langle v\rangle_-)/2$}, and represented by the dashed lines. These adiabatic limit values were evaluated from the velocities $\langle v\rangle_+$ and $\langle v\rangle_-$ obtained by simulations with static forces of absolute values $F_0$ in the positive and negative direction, respectively.  The excellent agreement with the simulation results for a time-dependent drive proves the validity of the adiabatic approximation up to frequencies $\omega\leq 10$. Above this threshold, $\lvr$ decreases with $\omega$ as a power low $\lvr \sim \omega^a$ with an exponent $a\simeq-4/3$. Thus the highest rectification velocities are achieved for small driving frequencies. The decrease of the velocity with the frequency is expected since a finite net-drift can only occur when the particles pass several periods of the channel within a half-period of the oscillating drive. In fact, by equating the time required to pass one period of the channel $\tau_d= L/\lvr$ with the half period $\tau_\mathrm{p}/2=\pi/\omega$ of the driving, one obtains a simple estimate of the characteristic frequency $\omega \sim \pi  \lvr$ beyond which the rectification velocity starts to decrease. This simple prediction agrees very well with the results of Fig.~\ref{v_over_freq}.

Since slow drivings provide the highest velocities, in the following a frequency of $\omega=\pi/10$ will be chosen. This frequency is high enough to allow a fast separation in a short time, while still low enough to be in the adiabatic regime. 

\section*{V. Improving Size-Dependent Particle Separation with the Entropic Splitter}\label{Sec_Separation}

\begin{figure}
\begin{center}
\includegraphics[width=8cm]{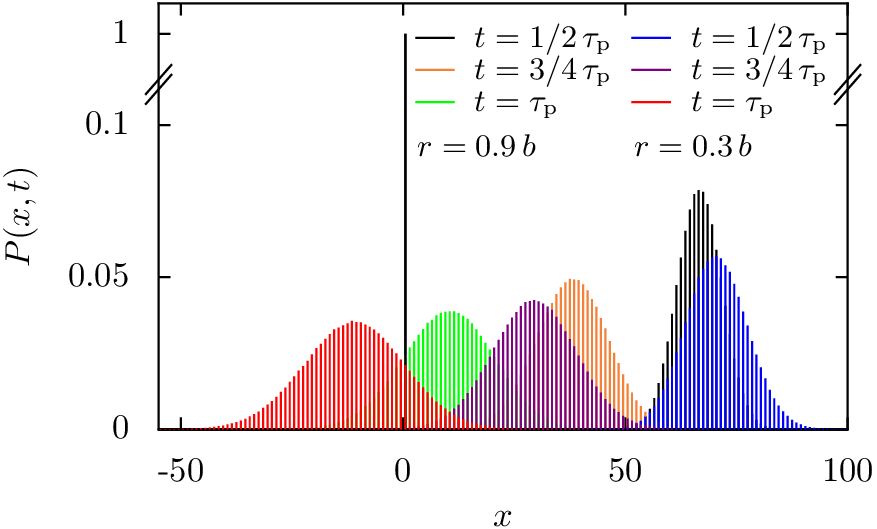}
\end{center}
\caption{(color online) Snapshots of the discretized marginal particle distribution $P(x,t)$ at different times $t$ for particles with radius $r=0.9\,b$ (left column in the legend) and \mbox{$r=0.3\,b$} (right column) with an amplitude $F_0=20$ of the time-periodic force $F(t)$ and a static bias $f_0=-7$. The initial distribution $P(x,t=0)$ is a uniform distribution in the range $0\leq x\leq L$, shown by the spike at $x=L/2$. One should note the higher negative velocity for $t>\tau_\mathrm{p}/2$ of the particles with $r=0.3\,b$. This leads to a negative net-drift of these particles after one period $\tau_\mathrm{p}$ whereas the particles with \mbox{$r=0.9\,b$} experience a positive net-drift.}
\label{Hist_x}
\end{figure}

In order to set up a system where two kinds of particles move in opposite directions, one has to apply an additional static bias~$f_0$ pointing in the opposite direction to the net rectification, i.e. in the negative direction. With this additional bias, the mean of the external forces is negative but due to the rectified motion particles with a radius above a certain threshold have a net-drift to the right, whereas particles below this threshold move to the left. To illustrate this mechanism of separation in opposite directions, Fig.~\ref{Hist_x} shows the marginal particle distribution given by \mbox{$P(x,t)=\left(1/\mathcal{P}\right)\int_{\loweref(x)}^{\upperef(x)} P(x,y,t)\,\mathrm{d}y$} where \mbox{$\mathcal{P}=\int_{0}^{L}\int_{\loweref(x)}^{\upperef(x)}P(x,y,t=0)\,\mathrm{d}y\,\mathrm{d}x$}. $P(x,t)$ is plotted for different times within one period of $F(t)$ with an amplitude~\mbox{$F_0=20$} and a static force $f_0=-7$, and for two different particle radii~$r=0.9b$ and $r=0.3b$.  One can see that the particles with $r=0.3b$ move faster to the right within the first half-period where $F(t)>0$ holds, whereas in the second half-period, when $F(t)<0$, the motion of the larger particles is disturbed more intensively caused by the smaller probability of passing the bottleneck compared to the smaller particles. This leads to a positive net-drift for the large particles with $r=0.9b$ and a negative net-drift for particles with $r=0.3b$ after one period of $F(t)$. 

\begin{figure}[t]
\begin{overpic}[width=8.50cm]{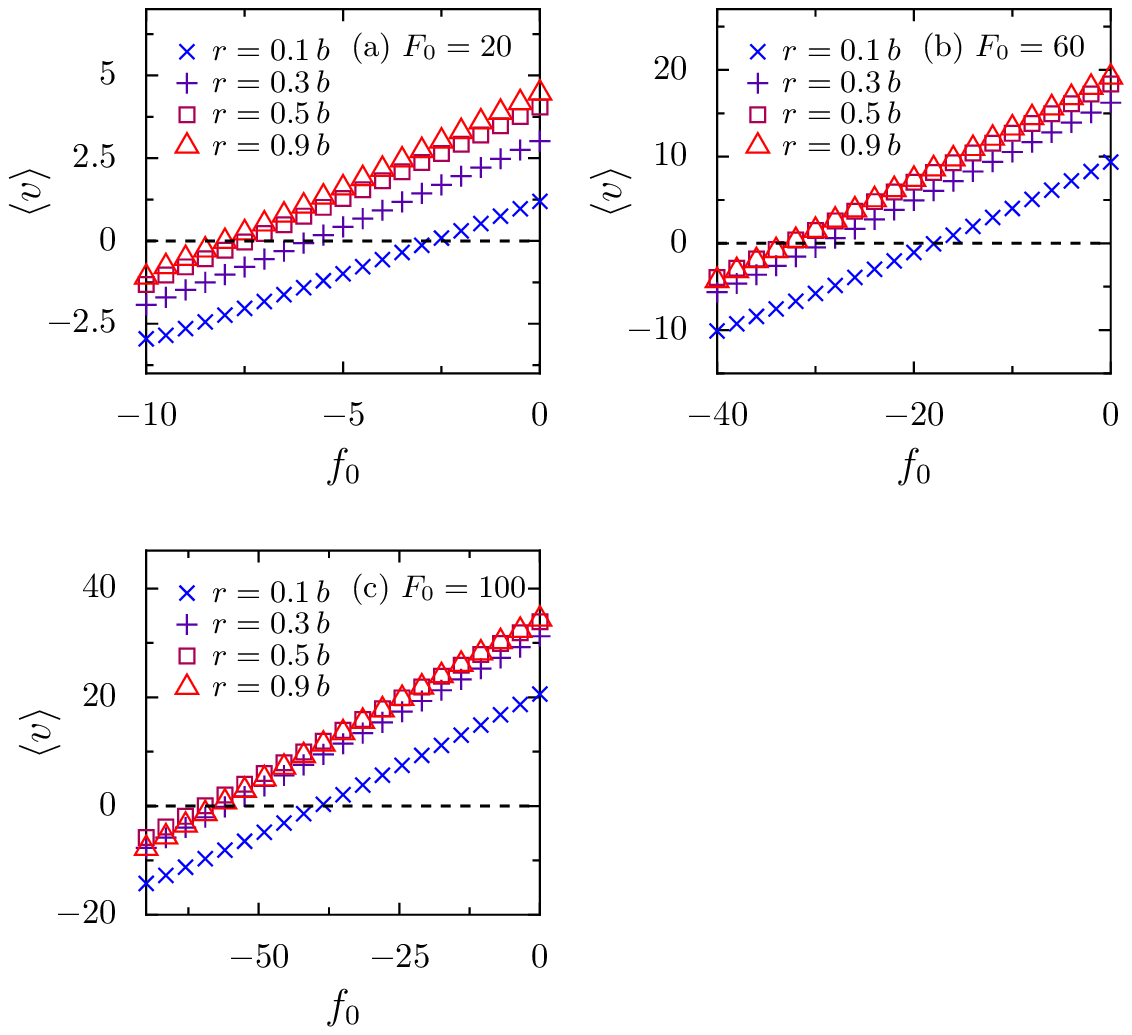}
\put(56,24){\begin{small}\begin{tabular}{C{0.6cm} C{0.6cm} C{0.6cm} C{0.6cm} | R{0.8cm}}\hline\hline
$r_\mathrm{s}/b$ & $r_\mathrm{l}/b$ & $F_0$ & $f_0$ & $\langle N_\mathrm{p}\rangle$ \\ \hline
$0.3$ & $0.9$ & $20$ & $-7$ & 63.2  \\ 
$0.1$ & $0.9$ & $20$ & $-5$ & 32.2 \\ 
$0.3$ & $0.9$ & $60$ & $-31$ & 20.5 \\
$0.1$ & $0.9$ & $60$ & $-25$ & 8.1 \\ 
$0.1$ & $0.9$ & $100$ & $-50$ & 4.0\\
$0.1$ & $0.5$ & $100$ & $-50$ & 4.0 \\ \hline\hline
\end{tabular}\end{small}}
\begin{tiny}\end{tiny}
\end{overpic}
\caption{The mean velocity $\lvr$ of a particle diffusing in the channel of Fig.~\ref{Geometrie} as a function of the static force $f_0$ for different values of the particle radii $r$ and the amplitude $F_0$ of the time-dependent force $F(t)$ (plots a-c). $\omega$ was chosen in the adiabatic regime with $\omega=\pi/10$. The table shows the mean number of oscillations $\langle N_p\rangle$ until all $2\times 10^4$ particles with the equally distributed radii $r_\mathrm{s}$ and $r_\mathrm{l}$ arrived the collectors at $x=\pm\,10^3 L$ of the device. It is shown that higher values of $F_0$ and $f_0$ deliver a faster splitting process, manifested in the lower averaged number of necessary oscillations $\langle N_\mathrm{p}\rangle$. Remarkably is the purity of 100\% that was obtained in every test shown in the table.}
\label{splitternew_fstat}
\end{figure}

A systematic analysis of the average velocities $\lvr$ for varying particle sizes and different combinations of the oscillating force's amplitude $F_0$ and a static bias~$f_0$ is shown in Fig.~\ref{splitternew_fstat}. The plots show a nearly linear behavior of $\lvr$ with respect to $f_0$. Thus, with the static bias one has a simple way to control the directionality of the motion of particles of a given size to achieve an efficient splitting. Since for  $F_0<100$ the rectification increases with $F_0$, both the values of the velocities and the strength of the static bias required to invert particle motion increase. However, this also leads to a decreasing difference of the mean velocities for different radii $r$ as $f_0$ becomes more negative. In particular, the plot with $F_0=100$ shows that only the velocities of small particles are really distinctive. Therefore, a separation of large particles with this large $F_0$ requires to choose $f_0$ very carefully. On the other hand, one should also note the scale of $\lvr$ in this plot since $F_0=100$ leads to large values of $\lvr$, an order of magnitude larger than those reported in Ref.~ \cite{Schmid}. This allows a very fast splitting with the limitation that only particles with $r\ll b$ can be easily separated from larger ones. This behavior is related to the particle accumulation near the center of the channel for large negative forces discussed in Fig.~\ref{muoverf_R_vgl_app}. Consequently, amplitudes~$F_0 < 10^2$ are most constructive for particle separation. If the size difference between particles is large, relatively large forces can be used for a quick splitting. To sort out particles with very similar sizes with the same channel geometry, smaller amplitudes of $F_0$ and $f_0$ can be used.

The splitting behavior for different radii $r$ and combinations of the amplitude $F_0$ and the static bias $f_0$ will now be tested in a specific example. The channel is assumed to have a total length of $2\times 10^3\,L$ and we will use $10^4$ particles of two different sizes to evaluate the separation efficiency. The particles are uniformly distributed in the interval $0\leq x\leq L$ at time $t=0$. The oscillatory and static forces are then switched on and the splitting procedure is continued until all $10^4$ particles of each kind are either in the left end of the device at $x=-10^3 L$ or at the right end at $x=10^3 L$. This procedure is repeated 50 times to calculate an averaged number of necessary oscillations $\langle N_\mathrm{p}\rangle$ of the periodic force until all particles reached a position $|x|>10^3L$. The table in Fig.~\ref{splitternew_fstat} shows the results of the tests for different combinations of $r$, $F_0$ and $f_0$. As can be seen, the average number of necessary periods $\langle N_\mathrm{p}\rangle$ is the lowest for the combination with the largest forces, corresponding to  $F_0=10^2$ and $f_0=-50$, as expected from the previous discussion and from the plots (a)-(c) in Fig.~\ref{splitternew_fstat}. More importantly, in all the cases shown in the table, all the small particles ended up at the left end of the device and all the large particles exited through the right end, thus achieving perfect purity in the separation.

To have an estimate in real units of the typical times, dimensions, and forces involved in the previous example, Table \ref{rescaled quantities} shows the values of the characteristic parameters for Brownian particles in water moving in a channel with period length $L\sim 1\,\mu\mathrm{m}$ at room temperature.

\begin{table}[h!!]
\begin{tabular}{r*{2}{R{7.0cm}}}\hline\hline
parameter & characteristic value \\ \hline  
 $L$ &  $1\,\mu\mathrm{m}$  \\ 
 $D_b$ \cite{Cussler} & $ 2\times\,10^{-12}\,\mathrm{m}^2/\mathrm{s}$ \\
$\tau$ & $0.5\,\mathrm{s}$ \\ 
$L/\tau$ & $2\, \mu \mathrm{m/s}$ \\
$\kb T/L$ & $10^{-15}\,\mathrm{N}$ \\ \hline\hline

\end{tabular}
\caption{Characteristic values of the main parameters used as an example to evaluate the performance of the entropic splitter.}
\label{rescaled quantities}
\end{table}

Thus, with a scaled driving frequency $\omega=\pi/10$ corresponding to a period $\tau_\mathrm{p}\sim 10\,\mathrm{s}$, the splitting with $F_0=100$ shown in the table takes a time that is on the order of $40\,\mathrm{s}$ in a channel with total length $2 \,\mathrm{mm}$. In combination with the purity of 100\% that was obtained in every test this results emphasize the potential of the entropic splitter. Similar purities of separation in shorter times can be achieved in devices with smaller total lengths.

 A simple estimate of the number of particles that can be separated in a given time can be made using the previous example. Assuming that the particles which have to be separated are continuously inserted into the device at $x=0$, a nearly average uniform distribution of the particles in the channel along the \mbox{$x$-axis} will be eventually reached, since the injected particles will compensate the loss of big and small particles that exit through different ends of the device. In order to be able to safely neglect particle-particle interactions, let us assume that at most one particle is situated in one period of the channel. This leads to an average total number of particles in the channel equal to the number of periods, yielding in our previous example a rate of 2000 separated particles within $40\,\mathrm{s}$. 
This is just an order of magnitude estimate of typical separation rates, since they obviously depend on the magnitude and frequency of the applied forces, the channel geometry, and the radii of the particles.
But, for practical use in massive separation, it is very important to note that it is technically possible to install many periodic channels in parallel. For example Kettner \textit{et al.}~\cite{Kettner} created a wafer pierced by approx. $10^6$ pores. This would lead to separation rates on the order of $2\times 10^9$ particles within $40\,\mathrm{s}$. Even using just a 2D parallelization, where all channels of the entropic splitter are placed in a single plane, this leads to a number of $10^3$ channels in a plane chip of 1-2 cm edge length and a rate of $2\times 10^6$ particles separated within $40\,\mathrm{s}$. Thus, very high purities at relatively high separation rates are feasible using many channels in parallel.

The plausibility of the perfect purity we obtained in the simulations can be supported by a simple analytic approximation. The separation purity can be estimated by the probability that a large particle with an average positive velocity reaches the ``wrong'' end of the device, i.e., the left
collector placed at $x=-nL$, where $n$ is the number of repeating periods of the channel~\cite{Schmid}. Assuming Gaussian distributed particles as it is suggested in Fig.~\ref{Hist_x}, the probability of finding a particle traveling at an average velocity $\lvr$ after the time $t$ at $x=-nL$ can be estimated by \mbox{$P(x<-nL,t)=0.5-0.5\mathrm{erf}\big[(nL+\lvr t)/\sqrt{4D_\mathrm{eff}t}\big]$}, where $D_\mathrm{eff}$  is the effective diffusion coefficient. $P(x,t)$ has a maximum at \mbox{$t_\mathrm{max}=nL/\lvr$}, which in scaled units is given by 
\mbox{$P(x<-n,t=t_{\mathrm{max}})=0.5-0.5\mathrm{erf}\big(\sqrt{n\lvr/D_\mathrm{eff}}\big)$}. 
Assuming $D_\mathrm{eff}\sim 1$, and a typical scaled velocity of $\lvr\sim 2$, a purity of $99.9996\%$ can be achieved already with a channel-length of only $10L$. 
With a very large total channel length of $2\times10^3 L$, the achieved purity is for all practical purposes of essentially $100\%$, as found in our test. Moreover, the previous calculation indicates that much shorter channels could have been used leading to even faster separation times and similar nearly perfect purities.

\section*{VI. Conclusions}

In summary, we examined how different parameters affect the sorting purity of the entropic splitter. We have found that it is possible to improve the entropic splitter presented by Reguera \textit{et al.} in Ref.~\cite{Schmid} using a slightly different geometry and by investigating a wide range of external forces acting on the particles. In particular, a maximum in the rectification efficiency can be found for amplitudes of the periodic forcing $F_0<10^2$. That leads to fast and efficient splitting of dissimilar particles. The application of even larger forces is not convenient, since the particles tend to be focused through the middle of the channel, thus not feeling the confining walls and loosing the rectification effect. In terms of frequencies, relatively small frequencies in the adiabatic regime offer the highest efficiencies. Once again, very high frequencies are not convenient since the particles would not have enough time to cross the bottlenecks and get rectified.
We verified the efficiency of splitting using specific tests in a wide range of forces, showing that a very fast splitting-performance with a high purity can be easily achieved. Hence, this improved configuration of the entropic splitter has the potential to be implemented in experiments and become a practical and efficient device for particle sorting.

\begin{acknowledgments}
The work was partially supported by the Spanish MINECO through Grant No. FIS2011-22603.
\end{acknowledgments}

\bibliography{literature_paper_splitter}

\end{document}